%% file: main.tex
\documentclass[sigconf]{acmart}

\usepackage{booktabs}
\usepackage{graphicx}
\usepackage{multirow}
\usepackage{subfig}
\usepackage{amsmath}
\usepackage{algorithm}
\usepackage{algpseudocode}
\usepackage{bookmark}

\AtBeginDocument{%
  }

\setcopyright{none}
\acmDOI{10.13140/RG.2.2.17165.76003}

\acmConference[FACCTREC'24]{The 7th FAccTRec Workshop: Responsible Recommendation}{Oct 14, 2024}{Bari, Italy}
\acmISBN{}

\begin{document}

% \title{The Name of the Title Is Hope}
\title{The trade-off between data minimization and fairness in collaborative filtering}

\author{Nasim Sonboli}
\email{nasim_sonboli@Brown.edu}
\orcid{0000-0002-6988-7397}
\affiliation{%
  \institution{Brown University}
  \city{Providence}
  \state{RI}
  \country{USA}
}

\author{Sipei Li}
\email{sipei.li@tufts.edu}
\affiliation{%
  \institution{Tufts University}
  \city{Medford}
  \state{MA}
  \country{USA}
}

\author{Mehdi Elahi}
 \email{mehdi.elahi@uib.no}
\affiliation{%
  \institution{University of Bergen}
  \city{Bergen}
  \country{Norway}
}

\author{Asia Biega}
\email{asia.biega@mpi-sp.org}
\affiliation{%
  \institution{Max-Planck Institute for Security and Privacy}
  \city{Bochum}
  \country{Germany}}

\renewcommand{\shortauthors}{Sonboli et al.}

\begin{abstract}
    General Data Protection Regulations (GDPR) aim to safeguard individuals' personal information from harm. While full compliance is mandatory in the European Union and the California Privacy Rights Act (CPRA), it is not in other places. GDPR requires simultaneous compliance with all the principles such as fairness, accuracy, and data minimization. However, it overlooks the potential contradictions within its principles. This matter gets even more complex when compliance is required from decision-making systems. Therefore, it is essential to investigate the feasibility of simultaneously achieving the goals of GDPR and machine learning, and the potential tradeoffs that might be forced upon us. This paper studies the relationship between the principles of data minimization and fairness in recommender systems. 

    We operationalize data minimization via active learning (AL) because, unlike many other methods, it can preserve a high accuracy while allowing for strategic data collection, hence minimizing the amount of data collection. We have implemented several active learning strategies (personalized and non-personalized) and conducted a comparative analysis focusing on accuracy and fairness on two publicly available datasets. The results demonstrate that different AL strategies may have different impacts on the accuracy of recommender systems with nearly all strategies negatively impacting fairness. There has been no to very limited work on the trade-off between data minimization and fairness, the pros and cons of active learning methods as tools for implementing data minimization, and the potential impacts of AL on fairness. By exploring these critical aspects, we offer valuable insights for developing recommender systems that are GDPR compliant.
\end{abstract}

% \ccsdesc[500]{Information systems~Relevance assessment}
% \ccsdesc[500]{Social and professional topics~User characteristics}
% \ccsdesc[500]{Applied computing~Law}


\maketitle

\section{Introduction}
    The European Union's General Data Protection Regulation \cite{EPRS} and CPRA (California Privacy Rights Act) \cite{ccpa_2018,cpra_2020} are both regulatory frameworks at the forefront of a global movement to strengthen individual privacy rights. They also impose significant fines for non-compliance. The EU can impose fines of up to \texteuro 20 million (US\$22 million) or 4\% of a company's global revenue, while the CPRA fines of up to \$7,500 per individual violation. As concerns about data privacy rise worldwide \cite{us_concern,us_privacy,presthus2018consumers}, we can expect more countries to enact similar legislation, creating a more consistent global landscape for data protection.

    Data protection regulations establish a framework of principles such as data minimization, accuracy, transparency, purpose limitation, storage limitation, etc. While compliance with all of these principles is mandatory, due to their conflicting goals, it might not be possible. Therefore, it is important to investigate the feasibility of simultaneously achieving these principles and learn the trade-offs we need to make, especially when we intend to build GDPR compliant decision making systems.

    % Similarly, Section 1798.100(c) of CPRA \cite{cpra_2020} stipulates that ``[...] collection, use, retention, and sharing of a consumer's personal information shall be reasonably necessary and proportionate to achieve the purpose for which the data was collected [...].'' 
    Here, we study the relationship between the principles of data minimization and fairness. Data minimization in Article 5(1)(c) of the European GDPR \cite{art5cgdpr} is defined as the responsibility to collect ``[...] adequate, relevant, and limited amount of personal data in relation to the purpose for which they are processed''. This principle limits the data collection and processing to what is necessary to achieve a pre-specified purpose. Non-compliance with data minimization leads to legal penalties \cite{dmrulings}.
    
    On the one hand, unlawful and excessive collection and misuse of data can lead to significant harm for users such as data breaches, surveillance, targeted censorship, filter bubbles, etc. On the other hand, under-collection, lack, or imbalance in the collected data, can equally exacerbate inequalities, especially for under-served groups. The fairness principle is intended to prevent such issues by requiring that the collection and processing of personal data must always be fair. Here we study this conflict. 

    Some studies have explored the trade-off between data minimization and accuracy and the feasibility of integrating this principle in machine learning algorithms \cite{Biega_DM_2020,finck2021reviving,rastegarpanah2021auditing,goldsteen2021_knowlg_Distll,Shanmugam2022_ScalingLaw}. However, the impact of data minimization on fairness remains largely unexplored.

    % This principle requests for responsible and ethical data processing. 
    For personalization systems, it is necessary to have enough data to deliver high-quality services. To operationalize data minimization in these systems, we choose to use Active Learning (AL), because of their ability to actively collect relevant data to achieve high accuracy. These methods weren't initially designed for data minimization, however, their strategy to selectively (and not overly) collect data and their priority to reach high accuracy with that data aligns with the goals of both data minimization and recommender systems. Other algorithmic techniques with similar properties can be potentially used for data minimization implementation, however they are outside the scope of this project. Later, we demonstrate how minimizing data via Active Learning leads to unfairness (accuracy imbalance) for the minority group in collaborative filtering systems.
    
    The main contribution of this article is listed below:
    \begin{itemize}
        \item Adopting active learning strategies as techniques for data minimization implementation in recommender systems for the first time, studying their pros and cons for data minimization to provide insight for future use. We implemented several well-established active learning strategies (personalized and non-personalized) on top of the most commonly used collaborative filtering method, matrix factorization (SVD), and assessed their performance. Our findings revealed that different active learning strategies may influence recommender systems' accuracy differently. 
        \item Investigating the impact of different data minimization strategies on the fairness of recommender systems. Our extensive offline experimentation of active learning strategies offers insights into the trade-offs between data minimization and fairness. Our results demonstrate that nearly all active learning strategies negatively impact the fairness of the output leading to disparate impact on the minority group. 
        \item Extending the literature on data minimization. The analysis from our empirical study provides insight for machine learning practitioners and researchers to adopt active learning strategies as techniques for data minimization implementation in recommender systems, considering the implications of such strategies from accuracy and fairness perspectives.
        % make a better decision about choosing the most appropriate technique for data minimization implementation for the recommendation algorithm that has the best trade-off with fairness and accuracy.
        \item Extending the current literature on active learning by comparing them through the lens of fairness, addresses the existing research gap. Numerous research efforts have investigated active learning strategies in recommender systems, but very limited work studied these strategies from a fairness standpoint.
    \end{itemize}

\section{Related Work}
\subsection{Data minimization}

        Operationalizing data minimization in machine learning requires overcoming several major hurdles including challenges in interpreting the law, a lack of clarity in the definition of the principle itself, the absence of a standard mathematical formulation, and a scarcity of technical guidance for organizations to follow \cite{finck2021reviving}. Despite these challenges, there is a growing interest in building methods that comply with the data minimization principle. This section summarizes the most common interpretation of data minimization for machine learning and proposed methods to implement data minimization in personalization.
        
        The core requirements of data minimization are adequacy, relevance, and purpose limitation. We interpret that the adequacy requirement refers to the \emph{amount} of data and relevance refers to its \emph{quality}. Quality should be defined with respect to the purpose. For instance, quality data for an ML system is the data that increases the accuracy of the model. Data that lacks either relevance or adequacy can prevent the system from completing its task. And without specifying the purpose, any data collection and processing is boundless. Therefore, data minimization allows the collection and processing of enough quality data to deliver a service but limits these actions to the purpose. The presented work here seeks to find a balance between minimizing the amount of data (adequacy) and increasing (or maintaining) the accuracy of their model (relevance).

    Biega at. al \cite{Biega_DM_2020} is the first article to study the feasibility and integration of data minimization in recommender systems. They provide two performance-based definitions for data minimization: global and per-user approaches. Both aim to minimize data per-user but with different approaches: \textit{global} focuses on maintaining average performance across users, while \textit{per-user} ensures each user meets a minimum performance threshold. Using their approach, they show the possibility of achieving quality recommendations with less data. They demonstrate empirically that minimization impacts individuals differently, potentially harming under-served groups with higher accuracy losses.

    Shanmugan et al. \cite{Shanmugam2022_ScalingLaw} propose a framework (FIDO) for recommender systems that uses the algorithm's performance curve for automatically determining and enforcing accurate stopping criteria for the data collection during training. This framework uses different active feature acquisition strategies. They demonstrate that accumulating more data doesn't always increase the per-user accuracy. If the collected data is not representative or is disparate, the data collection can hurt user performance. 
    % Here, we investigate this problem further for active learning methods.
 
    Clavell et al. \cite{Clavell_2020_AIES} investigate the tension between data minimization, performance, and fairness. Using a qualitative methodology, the authors conduct an algorithmic audit on REM{^^21}X (a health recommendation app) to identify and address the biases against the protected groups while adhering to GDPR's data minimization principle by avoiding the collection of sensitive user information. While they show they can minimize data while maintaining accuracy, they couldn't conduct bias assessments due to a lack of sensitive information. Therefore, collecting personal information becomes essential if its absence results in inaccuracies, or unfairness, or if the data is required for auditing and accountability purposes. Thus, data minimization should not be applied unless other legal principles of GDPR such as fairness are considered.

    Finally, Finck and Biega \cite{finck2021reviving} investigate if GDPR's data minimization and purpose limitation can be meaningfully implemented in data-driven systems with a focus on personalization and profiling. In their techno-legal analysis, they highlight the challenges and trade-offs of data protection laws, offer action points for the involved stakeholders, and emphasize the need for research on fairness-data minimization tension.  
    It's worth mentioning some of the work in data minimization is proposed for classification problems~\cite{rastegarpanah2021auditing,goldsteen2021_knowlg_Distll} and is out of the scope of this article. All of the previously mentioned articles emphasize the tension between data minimization and fairness and its importance. Here, we address this research gap.

    \subsection{Active learning}

Machine learning necessitates computational methods that capitalize on data to enhance system performance, often requiring substantial quantities of high-quality data \cite{bishop06pattern,Settles12ALbook,Olsson09survey}. Data acquisition can be resource-intensive, \sloppy time-consuming, and potentially unproductive, as it may include processing data that does not contribute significantly to the system's improvement. Active learning addresses these issues by strategic data selection, optimizing learning, and promoting data minimization.

Active learning encompasses a guided sampling process, wherein a system queries specific instances based on the data it has encountered thus far~\cite{Tong01Active}. Unlike passive learning, which relies on a predetermined dataset, active learning allows the learner to query, receive new data, and update its model accordingly. This selective approach to acquiring training data points reduces the number needed for model training ~\cite{Settles2010Active,rubens2015active} compared to passive learning enabling more efficient data collection.

Active learning in recommender systems is a crucial component that has been initially driven by the desire to create more efficient sign-up processes~\cite{desrosiersK11}. Using active learning, a recommender system actively selects items \cite{Bernard01improvingcollaborative,Teixeira2002ActiveCP,mehdi11system-wise,Sindhu2009Recom} or groups of items \cite{chang2015using,loepp2014choice} for users to rate. Evaluating the entire item set, the system identifies and presents items predicted to yield the most significant improvement in overall accuracy, enabling it to refine its recommendations based on user feedback.

Active learning strategies are a great fit to operationalize data minimization as they balance the goals of recommender systems and data minimization. Active learning strategies strive to increase or maintain the original accuracy of an algorithm while selectively collecting data. Both data minimization and active learning limit data collection and usage to a specific purpose. However, data minimization's priority is users' data protection, while active learning prioritizes accuracy. By using active learning techniques to implement data minimization, we strike a balance between system performance and prioritizing users' data rights.

\section{Methodology}

We represent a rating dataset consisting of $n$ users and $m$ items as a matrix $R = \{r_{ui}\} \in \mathbb{R}^{n \times m}$, where $r_{ui}$ denotes the rating provided by user $u$ to item $i$. $R$ could contain null-value entries when a user has never interacted with an item or has not yet provided a rating for that item. Indeed, the matrix $R$ is often sparse with large datasets such as the MovieLens 1M dataset~\cite{harper2015movielens}. 

An active learning strategy is a function $S(u, q, K, I_u) = L$, where $L = \{i_i, ..., i_{q'}\}, q' \leq q$, is a returned list of $q'$ selected items whose ratings should be queried from the user $u$, and $q$ is a hyper-parameter that represents the maximum number of items that the strategy could return. $K = \{k_{ui}\}\in \mathbb{R}^{n \times m}$ is a matrix of known ratings, i.e., $k_{ui}$ is a rating that has been acquired by the strategy from user $u$ on item $i$. $I_u$ is the set of items whose ratings have not been queried from the user $u$. The list of selected items $L$ is enforced to be a subset of $I_u$, i.e., $L \subset I_u$, and the active learning strategy will not query a user about the same item twice, meaning that once an item is selected by the strategy for user $u$ to rate, it will be removed from $I_u$ (more details in Algorithm \ref{alg:test}).

The known rating matrix $K$ is initially very sparse since the system starts with knowing little about the preferences of each user. In each querying step, the active learning strategy scores the items in $I_u$ for each user $u$, by utilizing the available information in $K$. When it is possible to score at least $q$ different items in $I_u$, the $q$ items with the highest scores are returned in $L$. Otherwise, all the $q' \leq q$ scored items are returned in $L$. Then user $u$ will provide ratings for the items in $L$ that they have interacted with before, which are added to $K$ so that the strategy will have more information for the next querying step. Note that user $u$ may have not experienced any of the items in $L$, leading to a possibility that the number of provided ratings is 0 and $K$ will not grow. Therefore, there could be a huge discrepancy between the number of ratings acquired by different active learning strategies. On the other hand, one could also argue that although querying about popular items will lead to a larger number of ratings, the information carried by those ratings is less than the ratings on rare items. This trade-off between rating quantity and rating quality is what sets different active learning strategies apart.

Common active learning strategies can be classified into three categories: uncertainty-reduction, error-reduction, and attention-based strategies. For our experimentation, following the experiments in \cite{Elahi2014_al}, we chose to implement at least one strategy from each category. Below, we describe them briefly.

% \todo[inline]{Sipei, please add the notations here and a pseudocode for how e implemented active learning}

{\bf Variance} strategy \cite{Bernard01improvingcollaborative,Teixeira2002ActiveCP} is an {\it uncertainty-reduction} method, focusing on selecting items with controversial or diverse ratings, as the system is more uncertain about users' opinions on these items \cite{Bernard01improvingcollaborative,Teixeira2002ActiveCP,Craig03Active}. One such strategy is the Variance strategy \cite{Bernard01improvingcollaborative,Teixeira2002ActiveCP}, which selects items with the highest variance in ratings, assuming that the variance reflects the system's uncertainty about the item's ratings. Items in $I_u$ with the highest rating variance in $K$ are selected until the desired size of $L$ is reached. 

% {\bf Greedy Extend} strategy is an {\it error reduction based} method aiming to reduce the system error by selecting items that help improve predictive accuracy \cite{Golbandi10Bootstraping, liu2011wisdom}. This is achieved by computing the reduction of RMSE for each candidate item in $I_u$, and selecting the item with the largest RMSE reduction until the desired size of $L$ is reached.
{\bf Greedy Extend} strategy is among {\it error reduction based} category of strategies that aim to reduce the system error by selecting items that help improve predictive accuracy the most\cite{Golbandi10Bootstraping,liu2011wisdom}. Greedy Extend \cite{Golbandi10Bootstraping} is one such strategy, which identifies items with ratings that, if added to $K$, will yield the lowest system RMSE \cite{Golbandi10Bootstraping,Golbandi11Adaptive}. This is achieved by computing the reduction of RMSE of the base recommender system when only including in $K$ each candidate item in $I_u$, and selecting the item with the largest RMSE reduction until the desired size of $L$ is reached. The calculation of RMSE will be discussed in more detail later.

{\bf Popularity} is an {\it attention-based} method that focuses on selecting items that have received the most attention among users, as they are more likely to be known and rated by users \cite{Teixeira2002ActiveCP,Carenini2003Toward,He11active}. Items in $I_u$ with the largest number of known ratings in $K$ are selected. While Popularity strategy \cite{rashid2002getting,Teixeira2002ActiveCP} is easy to implement and often serves as the baseline strategy \cite{Mello10Active,He11active,Golbandi10Bootstraping,Karimi11nonmyopic}, it may lead to prefix bias and may not provide much information to the system.

{\bf Popularity*Variance} is a {\it hybrid} method that aims to combine various individual strategies and balance the scores computed by each strategy. By multiplying the logarithm of the popularity score with the variance score of each item in $I_u$, {Popularity*Variance} make a balance of the quantity and quality of acquired ratings \cite{Golbandi10Bootstraping}. Accordingly, the effects of popularity and variance make a balance of the quantity and quality of acquired ratings. The use of a logarithm helps to transform the exponential distribution of popularity scores and reduces the weight of popularity. A variation of this strategy uses log(popularity) instead of $popularity$ and $entropy$ instead of $variance$  \cite{rashid2002getting}.

{\bf MaxRating} scores each item $i$ in $I_u$ using the rating prediction $\hat{r}_{ui}$ calculated by the base recommendation algorithm trained on the known ratings in $K$. Items with the highest predicted ratings are selected to be in $L$. MaxRating assumes that a higher predicted rating means a higher possibility that the user will rate this item and a higher possibility that this rating will provide useful information on the user's interests.

{\bf MinRating} uses the opposite heuristic as MaxRating, scoring each item $i$ using $\max \{r\} - \hat{r}_{ui}$, where $\max \{r\}$ is the maximum possible rating (e.g., 5). Therefore, MinRating selects items with the lowest rating predictions, gathering information on what users dislike, but are prone to getting less ratings.

{\bf MixedRating} is another example of the \textit{hybrid} category of strategies. MixedRating combines MaxRating and MinRating, selecting a mixture of items with the highest and lowest rating predictions $\hat{r}_{ui}$. Thus, MixedRating can gather information on what users like and dislike at the same time.

{\bf KNN} is a \textit{neighborhood-based} algorithm, which selects the top $L$ items in $I_u$ that are most similar to $K_u$, the items in the known set $K$ that user $u$ rated before. The item similarity is calculated as the cosine similarity of the rating arrays of two items.

{\bf Random} is a baseline strategy, which gives each item $i$ in $I_u$ a random score and selects the top $L$ items with the highest scores, hence a random selection.

\section{Experimental Setup}
    % \todo[inline]{which datasets? ML100k and ML 20M? and lastfm?}
    \textbf{Dataset.} We conduct experiments on MovieLens-1M~\cite{harper2015movielens}, a publicly available recommendation dataset. The dataset comprises integer ratings between 1 and 5 from 6,040 users on approximately 3,900 movies. We created a 5-core dataset (6,040 users, 3,377 items, density of \%8), where every user and item has at least 5 ratings. We chose women as the protected group due to their lower count and smaller profile sizes (4,331 men and 1,709 women). The \textit{userfixed} technique was used to split the data where 80\% of the ratings of every user are used as their training set and 20\% as their test set. Due to lack of space, only the results on MovieLens-1M are shown.
    % \todo[inline]{we need to add another dataset! do we know of any?}

% This\hspace{-0.2cm}is\hspace{-0.2cm}tight.

    \textbf{Recommendation Algorithms.} We chose matrix factorization (SVD)~\cite{funk2006netflix} as our base recommender system. We used the Surprise~\cite{hug2020surprise} implementation of the SVD and tuned the parameters using GridSearchCV from scikit-learn library~\cite{scikitlearn}. Finally, we ran SVD with 100 latent factors, a learning rate of 0.005, and a regularization term of 0.1 over 5-fold cross-validation.
    Let $r_{ui}$ be the rating from user $u$ to item $i$, and $\hat{r}_{ui}$ be the prediction of $r_{ui}$ by the recommendation algorithm. The rating prediction is calculated by matrix factorization as 
    % \vspace{-0.9cm}
        % \raggedbottom
        \begin{equation}
            \hat{r_{ui}} = \mu + b_u + b_i + q_i^T p_u
        % \vspace{-1.5cm}
        \end{equation}

    , where $\mu$ is the average rating of the training set, $b_u$ and $b_i$ are the corresponding bias terms of user $u$ and item $i$, and $p_u$ and $q_i$ the corresponding latent factors learned by the system.

    % \todo[inline]{cite \href{https://surpriselib.com/}{Surprise}}
    \textbf{Evaluation Metrics.} To evaluate the \textit{performance} of the recommendation algorithms, we use Root Mean Squared Error(RMSE) which calculates the difference between the true and predicted ratings of all the items in the test set. To measure the \textit{unfairness} or performance imbalance, we compare RMSE differences.

    % \todo[inline]{Sipei, could you please add the formula here}
    % \vspace{-0.5cm}
    % \begin{equation}
    %     RMSE = \sqrt{\frac{1}{|T|}\sum_{r_{ui}\in T}(r_{ui} - \hat{r_{ui}})^2}
    %     % \vspace{-0.9cm}
    % \end{equation}

    % \begin{small}
    \begin{equation*}
    % \begin{aligned}
        \text{RMSE} = \sqrt{\frac{1}{|T|}\sum_{r_{ui}\in T}(r_{ui} - \hat{r}_{ui})^2}
    % \end{aligned}
    \label{eq:rmse}
    \end{equation*}
    % \end{small}
    
    , where $T$ is the test set. When calculating the RMSE for female users, we evaluated on a subset of $T$, $T_f = \{r_{ui} \in T | u \text{ is female }\}$. Similarly, for male users we used $T_m = \{r_{ui} \in T | u \text{ is male }\}$.
    % \textbf{Data Minimization}
    % Information Loss? Are we minimizing quantitatively or qualitatively?
    % measure how the rating counts become lower after active learning algo while preserving accuracy
  
    % \subsection{Baselines}
    \textbf{Compared Methods.} We implemented and compared five personalized and four non-personalized active learning strategies (explained in the Methodology Section). 
    % The source code is available on our GitHub page.\footnote{}
    % \todo[inline]{should we put the al strategies that we used here?}
    
    % \textbf{Implementation Details}
    % 0.02\% known dataset 
    % a sliding window of 10
    % 120 iterations?25?
    % \todo[inline]{details of the implementation and how the AL data selection works in general, what is known set, candidate set, etc.}
    \begin{algorithm}[htbp]
        \caption{The testing algorithm for a strategy $S$}\label{alg:test}
        \begin{algorithmic}[1]
            \Require Dataset $R$, strategy $S$, base recommendation model $M$, gender mapping $G$, query size $q$
            \Ensure RMSE of female users after each query $RMSE_f$, RMSE of male users after each query
            \State $X, T \leftarrow $ userfixed\_split($R$).
            \State $K \leftarrow$ randomly sampled 0.2\% of $X$.
            \State $X \leftarrow X \setminus K$.
            \State $T_f, T_m \leftarrow$ $T$ partitioned based on $G$. 
            \For{each user $u$}
            \State $I_u \leftarrow \{i | k_{ui} = NULL\}$.
            \EndFor
            \State $RMSE_f \leftarrow $ empty list.
            \State $RMSE_m \leftarrow $ empty list.
            \While{$\exists I_u \neq \emptyset$}
            \For{each user $u$ s.t. $I_u \neq \emptyset$}
            \State $L \leftarrow S(u, q, K, I_u)$
            \State $L_e \leftarrow \{i \in L | x_{ui} \neq NULL\}$.
            \For{$i \in L_e$}
            \State $k_{ui} \leftarrow x_{ui}$. 
            \State $X \leftarrow X \setminus x_{ui}$
            \EndFor
            \State $I_u \leftarrow I_u \setminus L$.
            \EndFor
            \State Train $M$ on $K$.
            \State $RMSE_f.append(RMSE(T_f, M(T_f)))$
            \State $RMSE_m.append(RMSE(T_m, M(T_m)))$
            \EndWhile
        \end{algorithmic}
    \end{algorithm}
        
\section{Results and Discussion}

    \input{tables/ML/ml_proUnpro_unpers_rmse_table}
    \input{tables/ML/ml_proUnpro_pers_rmse_table}
    The training data is split into a known set $K$ (0.2\% of the training data) and a candidate set. At each iteration, $W=10$ ratings from the candidate set are added to the known data, based on a chosen active learning strategy, until all the candidate items are added. The recommendation algorithm is trained on $K$ and tested against the test set. The test set was kept the same for all the algorithms and iterations. MixedRating, MinRating, MaxRating, RandomP, and KNN are personalized strategies while Pop, Pop-Var, Var, Greedy Extend(GE), and Random are non-personalized.

    \subsection{Data Minimization through Active Learning}
    %Generally, the more data is added, the lower the RMSE gets for all of the strategies. 
    
\begin{figure}[tp!] % h = here, t = top, b = bottom, p = page
            \centering
            \includegraphics[width=1\linewidth]{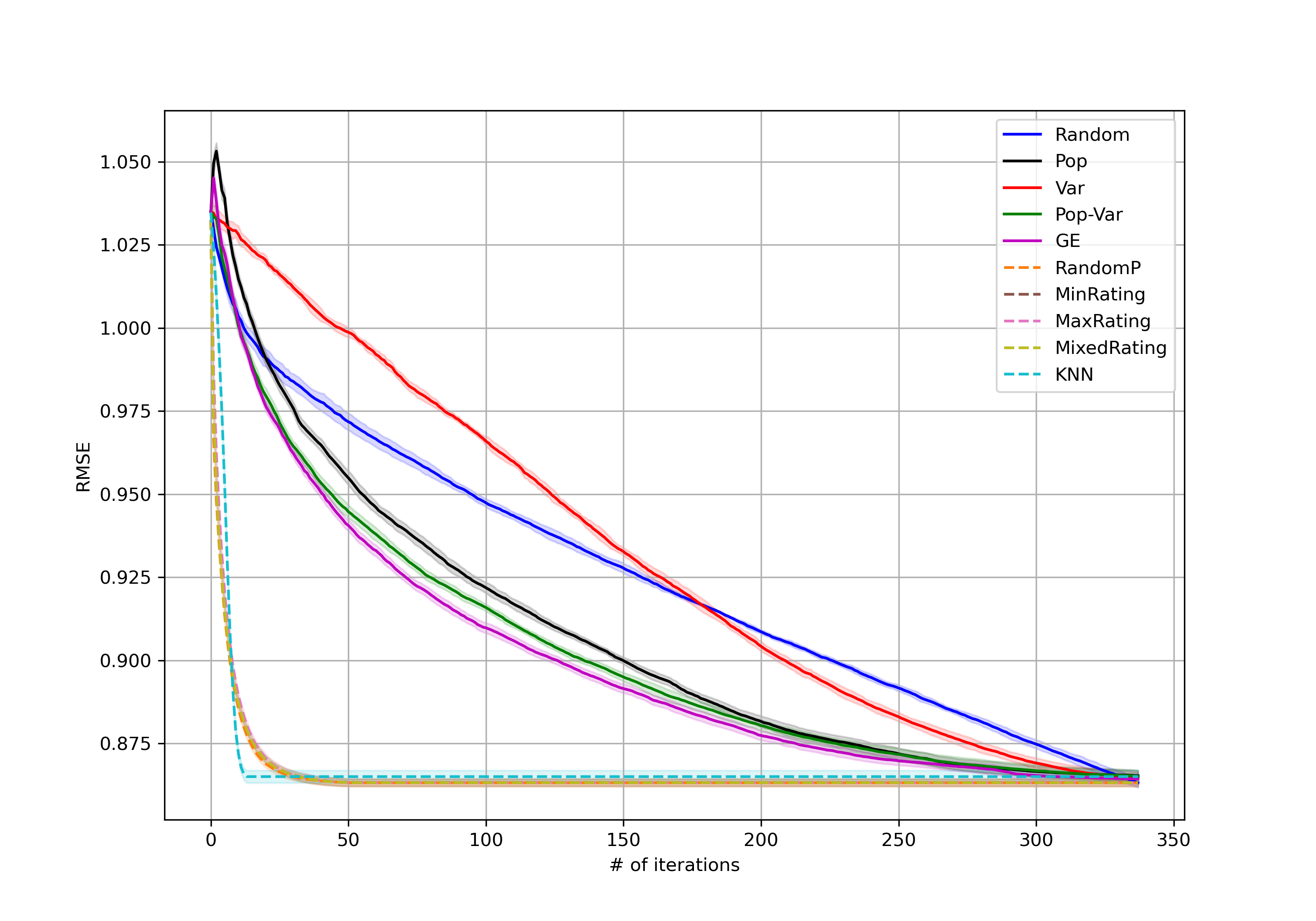}
            \caption{Experiment\#1: RMSE trend of personalized \& non-personalized active learning strategies in MovieLens dataset over 340 iterations, with a sliding window of 10 items}
            \Description{exp\#1 RMSE}
            \label{fig:exp1_rmse}
    \end{figure}

    Figure \ref{fig:exp1_rmse} shows the results of all the strategies on the MovieLens-1m dataset. The error starts from 1.05 and decreases to around 0.875. Personalized strategies show lower errors. There are no statistically significant differences between the RMSE of personalized methods except for Knn. They initially accumulate more data in the first 10 iterations and plateau quickly while the RMSE of the non-personalized strategies gradually decreases. Non-personalized strategies are best suited for eliciting data on new users or new items when personal data is not accessible or in contexts where the use of personal data is not allowed by the law.
    Var and Random show consistently higher RMSE compared to the rest of the algorithms and accumulate the least number of ratings. MovieLens dataset's innate popularity bias (having more data from the majority group on popular items) leads to an RMSE decrease for strategies that select more popular items such as Pop and Pop-var. Pop-Var shows a slightly lower RMSE than Pop. GE performs the best among the non-personalized methods.

    According to Table \ref{tab:pers} and Table \ref{tab:nonpers}, non-personalized strategies have lower RMSE differences (more fair). Pop and Pop-Var consecutively have the lowest RMSE difference up to iteration 160, where Random starts to outperform both. GE shows the lowest RMSE but the highest RMSE difference. Pop-Var has one of the lowest RMSEs and also the lowest RMSE differences, making it a good choice for balancing data minimization and fairness. Among the personalized methods, Knn initially shows a lower RMSE and lowest RMSE difference.

    Active Learning algorithms have different impacts on fairness. As it is seen in Table \ref{tab:pers}, the RMSE of the protected group is generally higher than that of the unprotected group. Table \ref{tab:nonpers} and Table \ref{tab:pers} demonstrate the calculated RMSE of both user groups (men and women) in all the AL strategies for the MovieLens dataset.

    Note that, the RMSE results might be influenced by the ratio of the protected and unprotected group ratings in the test set, in the known set, or the training. For example, suppose the initial ratio of the unprotected group is not the same as its ratio in the test set, or the data in the train set doesn't represent the patterns of the unprotected group as in the test set. In that case, their RMSE might be a lower number. If the ratio of the unprotected is higher than the protected group, it influences the overall RMSE more. This issue will affect the output RMSE for the protected group. In the next section, we will explore if an equal ratio of the protected and unprotected groups in the input will influence Fairness. The rest of the experiments are left for future work.

    %This issue requires additional analysis of the influence of the user group ratios on the results.

    \subsection{Equal Ratio Input}
    % % equal ratio experiments
    In this experiment, we collect the same number of ratings from men and women. The difference between the GE, Random, and Var is more distinctive up to iteration 200 when the user group ratio is balanced. It shows the influence of the data from the majority group on the overall RMSE. All the strategies show lower RMSE differences between the protected and unprotected groups. Similarly, non-personalized methods show lower RMSE difference than the personalized methods. In ML-100k the RMSE difference is proportionate to the ratio of the protected and unprotected. However, in ML-1m, the results reverse where the protected group reaches higher accuracy than the unprotected. This result should be further investigated.

    \begin{figure}[htbp]
            \centering
            \includegraphics[width=1\linewidth]{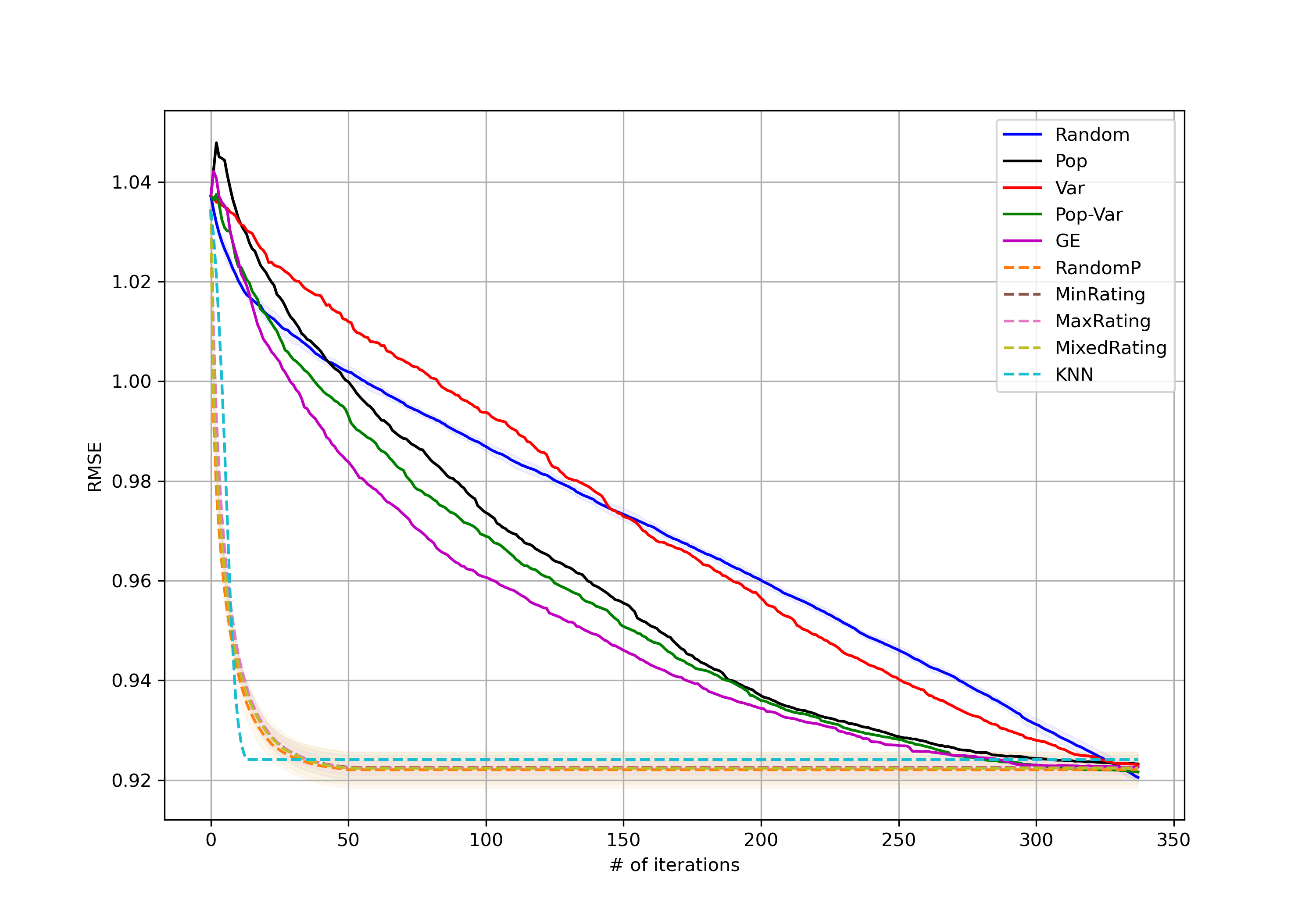}
            \caption{Experiment\#2: RMSE trend of personalized \& non-personalized active learning strategies in MovieLens dataset over 340 iterations, with a sliding window of 10 items}
            \Description{exp\#2 RMSE}
            \label{fig:exp2_rmse}
        \end{figure}

\section{Conclusion and Future Work}

    In this work, we investigate the relationship between fairness and active learning. We operationalized data minimization via active learning methods, explored the extent to which these methods reduce the amount of data and general RMSE, and demonstrated how this data loss affects the minority groups more. 

    Data minimization's main intention is not merely reducing the amount of data. Data minimization is about allowing data collection and usage after a lawful purpose is explicitly specified. This data should be relevant to the purpose and adequate to fulfill the purpose. Generally, data minimization prevents the unlawful over-collection and processing of data, the collection of irrelevant data to the purpose, and the collection for a secondary purpose that is not aligned with the initial purpose.
    
    We showed sometimes a more balanced accuracy result is possible with a better data representation. In our experiments, we under-sampled the majority group which led to a major loss of accuracy for that group. However, in reality, more information can be collected for minorities to close the RMSE gap between the groups. It means to reach a more balanced accuracy among all the user groups, sometimes we might need to collect more data for that group than minimize its amount.

    In the second experiment, we balanced the ratio of protected and unprotected in the collected data. These experiments showed that the amount of data and the ratio of it affects the results significantly. In one dataset, the RMSE gap decreased and in the other dataset, the results were reversed! This can mean that merely the amount of data might not be the most important factor in the decision-making of systems. Instead, the information that the data contributes to the model matters, and the adequacy of that information to make accurate inferences for different user groups.

    Data minimization through active learning widened the RMSE gap in the results which can lead to inequalities. Therefore, to design machine learning algorithms that comply with a data protection regulation such as data minimization, one needs to think about other GDPR regulations such as accuracy, fairness, etc. Investigating these conflicts is very important as simultaneous compliance is expected.

    Our results are limited to active learning techniques. To generalize these results, we intend to expand our experiments to other state-of-the-art recommendation algorithms. Although our results are limited to active learning methods, any methodology that samples or uses data disproportionately can potentially cause unfairness in accuracy. Therefore, we need to design algorithms that consider these conflicts. In the future, we need to test our hypothesis on other datasets and include other baseline algorithms such as KNN, or state-of-the-art algorithms such as neural networks. 

    Our analysis is limited to the RMSE metric in this paper to see the raw impact of data minimization on group ratings. However, we intend to extend this study by using more complex provider-side and consumer-side fairness metrics \cite{Sonboli2021_librecauto,deldjoo2021flexible} and ranking-based metrics.

\begin{acks}
Author Sonboli's effort was supported by the CRA CIFellows program funded by the National Science Foundation under Grant No.~2127309.
\end{acks}

\bibliographystyle{ACM-Reference-Format}
\bibliography{main}

\end{document}

%% file: tables/ML/ml_proUnpro_unpers_rmse_table.tex
\renewcommand{\arraystretch}{0.8}
\begin{table*}[htbp!]
\tiny
\setlength\tabcolsep{0.5pt}
\begin{tabular*}{\textwidth}{@{\extracolsep{\fill}} ccccccccc|ccccccc}
\toprule[1.1pt]
\multicolumn{1}{l}{}     &        &         \multicolumn{7}{c}{Original Ratio}            &            \multicolumn{7}{c}{Equal Ratio}          \\
\cmidrule(lr){3-9}
\cmidrule(lr){10-16}
                         &        & i=0    & i=50   & i=100  & i=150  & i=200  & i=250  & i=300  &  i=0   & i=50   & i=100  & i=150  & i=200  & i=250 & i=300 \\ 
\midrule
\multicolumn{16}{c}{RMSE}                                                                                                                         \\ 
\midrule
\multirow{2}{*}{Random}  & Female & 1.048* & 0.994* & 0.971* & 0.953* & 0.936* & 0.920* & 0.906* & 1.046* & 0.997* & 0.978* & 0.959* & 0.942* & 0.925* & 0.908* \\
                         & Male   & 1.031* & 0.965* & 0.940* & 0.919* & 0.900* & 0.882* & 0.864* & 1.034* & 1.004* & 0.990* & 0.978* & 0.966* & 0.953* & 0.939* \\ 
\midrule
\multirow{2}{*}{Pop}     & Female & 1.048* & 0.972* & 0.944* & 0.925* & 0.910* & 0.902* & 0.898* & 1.046* & 0.976* & 0.945* & 0.926* & 0.910* & 0.902* & 0.898* \\
                         & Male   & 1.031* & 0.949* & 0.914* & 0.892* & 0.872* & 0.862* & 0.856* & 1.034* & 1.008* & 0.983* & 0.965* & 0.946* & 0.937* & 0.933* \\ 
\midrule
\multirow{2}{*}{Var}     & Female & 1.048* & 1.025* & 0.995* & 0.963* & 0.935* & 0.914* & 0.900* & 1.046* & 1.024* & 1.001* & 0.970* & 0.944* & 0.920* & 0.903* \\
                         & Male   & 1.031* & 0.990* & 0.956* & 0.923* & 0.894* & 0.873* & 0.859* & 1.034* & 1.008* & 0.991* & 0.974* & 0.960* & 0.947* & 0.936* \\ 
\midrule
\multirow{2}{*}{Pop-Var} & Female & 1.048* & 0.965* & 0.939* & 0.921* & 0.909* & 0.902* & 0.898* & 1.046* & 0.970* & 0.940* & 0.923* & 0.909* & 0.903* & 0.898* \\
                         & Male   & 1.031* & 0.938* & 0.908* & 0.886* & 0.871* & 0.862* & 0.856* & 1.034* & 1.000* & 0.978* & 0.960* & 0.944* & 0.936* & 0.931* \\ 
\midrule
\multirow{2}{*}{GE}      & Female & 1.048* & 0.966* & 0.939* & 0.923* & 0.908* & 0.901* & 0.897* & 1.046* & 0.968* & 0.941* & 0.923* & 0.908* & 0.901* & 0.898* \\
                         & Male   & 1.031* & 0.932* & 0.900* & 0.881* & 0.867* & 0.859* & 0.855* & 1.034* & 0.989* & 0.967* & 0.953* & 0.943* & 0.935* & 0.931* \\ 
\bottomrule[1.1pt]
\end{tabular*}
\captionsetup{size=footnotesize}

\caption{Table 1: RMSE of non-personalized active learning strategies for female and male users. * denotes cases where the difference in RMSE between female and male users is statistically significant under a two-tailed t-test with $p < 0.01$.}
\label{tab:nonpers}
\end{table*}

%% file: tables/ML/ml_proUnpro_pers_rmse_table.tex
\renewcommand{\arraystretch}{0.8}
\begin{table*}[htbp!]
\tiny
\setlength\tabcolsep{0.5pt}
\begin{tabular*}{\textwidth}{@{\extracolsep{\fill}} ccccccccc|ccccccc}
\toprule[1.1pt]
\multicolumn{1}{l}{}         &        &         \multicolumn{7}{c}{Original Ratio}            &            \multicolumn{7}{c}{Equal Ratio}          \\
\cmidrule(lr){3-9}
\cmidrule(lr){10-16}
                             &        & i=0    & i=5    & i=10   & i=20   & i=30   & i=40   & i=50   &  i=0   & i=5    & i=10   & i=20   & i=30   & i=40   & i=50  \\ 
\midrule
\multicolumn{16}{c}{RMSE}                                                                                                                                      \\ 
\midrule
\multirow{2}{*}{RandomP}     & Female & 1.047* & 0.940* & 0.914* & 0.899* & 0.896* & 0.895* & 0.895* & 1.046* & 0.948* & 0.923* & 0.904* & 0.899* & 0.896* & 0.895* \\
                             & Male   & 1.028* & 0.905* & 0.876* & 0.859* & 0.855* & 0.853* & 0.853* & 1.031* & 0.961* & 0.947* & 0.936* & 0.933* & 0.931* & 0.931* \\ 
\midrule
\multirow{2}{*}{MinRating}   & Female & 1.047* & 0.943* & 0.917* & 0.901* & 0.897* & 0.895* & 0.895* & 1.046* & 0.951* & 0.925* & 0.906* & 0.900* & 0.897* & 0.896* \\
                             & Male   & 1.028* & 0.908* & 0.878* & 0.860* & 0.855* & 0.853* & 0.853* & 1.031* & 0.966* & 0.949* & 0.938* & 0.934* & 0.932* & 0.931* \\ 
\midrule
\multirow{2}{*}{MaxRating}   & Female & 1.047* & 0.945* & 0.918* & 0.901* & 0.896* & 0.895* & 0.895* & 1.046* & 0.952* & 0.925* & 0.906* & 0.899* & 0.896* & 0.895* \\
                             & Male   & 1.028* & 0.912* & 0.880* & 0.861* & 0.855* & 0.853* & 0.853* & 1.031* & 0.971* & 0.952* & 0.938* & 0.934* & 0.932* & 0.931* \\ 
\midrule
\multirow{2}{*}{MixedRating} & Female & 1.047* & 0.942* & 0.916* & 0.901* & 0.897* & 0.895* & 0.895* & 1.046* & 0.950* & 0.924* & 0.905* & 0.899* & 0.896* & 0.895* \\
                             & Male   & 1.028* & 0.907* & 0.877* & 0.860* & 0.855* & 0.853* & 0.853* & 1.031* & 0.966* & 0.949* & 0.938* & 0.934* & 0.932* & 0.931* \\ 
\midrule
\multirow{2}{*}{KNN}         & Female & 1.048 & 0.968* & 0.900* & 0.895* & 0.895* & 0.895* & 0.895* & 1.049* & 0.978* & 0.903* & 0.895* & 0.895 & 0.895* & 0.895* \\
                             & Male   & 1.030 & 0.948* & 0.863* & 0.855* & 0.855* & 0.855* & 0.855* & 1.029* & 0.990* & 0.940* & 0.934* & 0.934* & 0.934* & 0.934* \\ 
\bottomrule[1.1pt]
\end{tabular*}
\captionsetup{size=footnotesize}
\caption{Table 2: RMSE of personalized active learning strategies for female and male users. * denotes cases where the difference in RMSE between female and male users is statistically significant under a two-tailed t-test with $p < 0.01$.}
\label{tab:pers}
\end{table*}